\documentclass[aps,prb,twocolumn,reprint,showpacs,superscriptaddress]{revtex4-1}
\usepackage{amsmath}
\usepackage{amssymb}
\usepackage{graphicx}
\usepackage{hyperref}
\usepackage{color,xcolor}
\begin{document}
\title{Topological magnetic phase in LaMnO$_3$ ($111$) bilayer}
\author{Yakui Weng}
\author{Xin Huang}
\affiliation{Department of Physics and Jiangsu key laboratory for advanced metallic materials, Southeast University, Nanjing 211189, China}
\author{Yugui Yao}
\affiliation{Department of Physics, Beijing Institute of Technology, Beijing 100080, China}
\author{Shuai Dong}
\email{sdong@seu.edu.cn}
\affiliation{Department of Physics and Jiangsu key laboratory for advanced metallic materials, Southeast University, Nanjing 211189, China}
\date{\today}

\begin{abstract}
Candidates for correlated topological insulators, originated from the spin-orbit coupling as well as Hubbard type correlation, are expected in the ($111$) bilayer of perovskite-structural transition-metal oxides. Based on the first-principles calculation and tight-binding model, the electronic structure of a LaMnO$_3$ ($111$) bilayer sandwiched in LaScO$_3$ barriers has been investigated. For the ideal undistorted perovskite structure, the Fermi energy of LaMnO$_3$ ($111$) bilayer just stays at the Dirac point, rendering a semi-metal (graphene-like) which is also a half-metal (different from graphene nor previous studied LaNiO$_3$ ($111$) bilayer). The Dirac cone  can be opened by the spin-orbit coupling, giving rise to nontrivial topological bands corresponding to the (quantized) anomalous Hall effect. For the realistic orthorhombic distorted lattice, the Dirac point moves with increasing Hubbard repulsion (or equivalent Jahn-Teller distortion). Finally, a Mott gap opens, establishing a phase boundary between the Mott insulator and topological magnetic insulator. Our calculation finds that the gap opened by spin-orbit coupling is much smaller in the orthorhombic distorted lattice ($\sim$$1.7$ meV) than the undistorted one ($\sim$$11$ meV). Therefore, to suppress the lattice distortion can be helpful to enhance the robustness of topological phase in perovskite ($111$) bilayers.
\end{abstract}
\maketitle

\section{Introduction}
Heterostructures of transition-metal oxides (TMOs) have become one of the most attractive research topics in condensed matter physics and material science. On one hand, the wide classes of available materials of TMOs provide a multitude of intriguing physical properties, such as superconductivity, ferromagnetism, and ferroelectricity, which can be incorporated into nanoscale multilayers. On the other hand, the latest thin film technology can precisely control oxides growth up to atomic level, which offers the possibility of device design based on these correlated electronic materials.\cite{Dagotto:Sci07,Mannhart:Sci,Takagi:Sci,Hammerl:Sci}

In particular, the tailoring of structural orientation is an effective route to tune the electronic properties of oxide heterostructures. For example, orientation-dependent magnetism has been revealed in many oxide heterostructures, e.g. LaFeO$_3$/LaCrO$_3$ \cite{Ueda:Sci,Ueda:Jap,Zhu:Jap}and LaNiO$_3$/LaMnO$_3$.\cite{Gibert:Nm,Dong:Prb13} In addition, topological phases were predicted to emerge in the (111) bilayers of various perovskite-type TMOs,\cite{Xiao:Nc11} since a ($111$) bilayer of perovskite forms a buckled honeycomb lattice as in silicene.\cite{Liu:Prl} It is well known that the Dirac-cone-type band structure can be easily observed in such a honeycomb lattice, while it is not common in pseudocubic lattices. According to previous studies, topological phases usually emerge in these systems with Dirac-cone-type bands.\cite{Hasan:Rmp,Vafek:Arcmp}

Following this idea, the LaNiO$_3$ ($111$) bilayer sandwiched in LaAlO$_3$ barriers was studied, giving a topological insulating phase in a very narrow region driven by strong Coulomb interaction up to $U\sim6$ eV.\cite{Yang:Prb,Ruegg:Prb11,Ruegg:Prb12,Ruegg:Prb13,Doennig:Prb} However, it is well known that LaNiO$_3$ itself is a paramagnetic metal, implying a very weak Hubbard repulsion, otherwise a magnetic insulator would be obtained. Thus, such a prediction, although very novel in theory, may become nonrealistic and experimentally unaccessible. Therefore, to find a proper perovskite system becomes meaningful to pursuit correlated topological phase.

In this study, another well-known perovskite, LaMnO$_3$, will be studied in the form of ($111$) bilayer, to pursuit more realistic topological phase. For non-magnetic LaNiO$_3$, the $e_{\rm g}^1$ configuration is only quarter-filled for the $e_{\rm g}$ sector, leaving three unoccupied $e_{\rm g}$ orbitals.\cite{Yang:Prb,Ruegg:Prb11} While for magnetic LaMnO$_3$, due to the strong Hund coupling between the half-filled $t_{\rm 2g}$ electrons and $e_{\rm g}$ electron, the high-spin state makes the $e_{\rm g}^1$ configuration to be spin-polarized half-filling. Therefore, these two systems are quite different regarding the physical properties.

\begin{figure}
\centering
\includegraphics[width=0.5\textwidth]{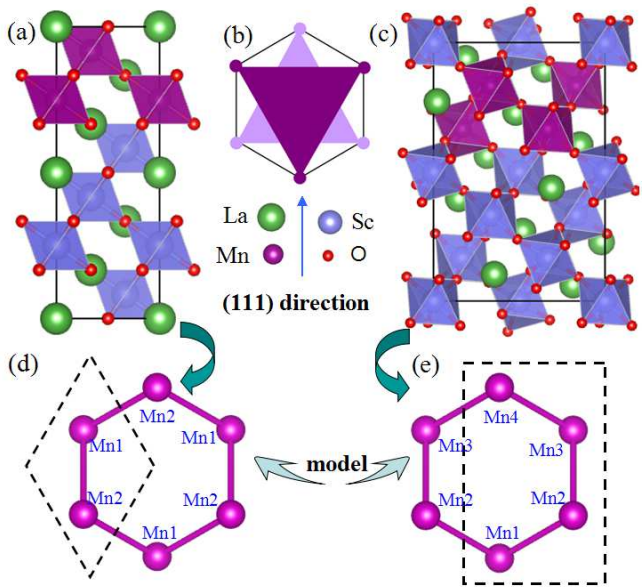}
\caption{(Color online) Superlattice structures of the (LaMnO$_3$)$_2$/(LaScO$_3$)$_4$ grown along the [$111$] direction: (a) A unit cell constructed by cubic perovskites (without distortions). (b) Top view of Mn's plane along the [$111$] direction. Two colors (purple and lilac) denote the top and bottom layers of Mn ions. (c) A unit cell constructed by orthorhombic perovskites (with distortions). (d-e) Sketches of Mn's honeycomb lattices. The unit cells are marked by black broken lines. The identical structural characters, A-site cations, and approximate lattice constants ensure the possibility for epitaxial growth of LaMnO$_3$/LaScO$_3$ heterostructure in experiments.
}
\label{F1}
\end{figure}

In following, the (LaMnO$_3$)$_2$/(LaScO$_3$)$_4$ superlattice, as sketched in Fig.~\ref{F1}, will be studied using density-functional theory (DFT) as well as tight-binding model. LaScO$_3$ is a nonmagnetic band insulator with a large band gap up to $6.0$ eV.\cite{Afanas:Apl,He:Prb12} Thus, the LaMnO$_3$ bilayer is perfectly isolated to form a two-dimensional lattice, as required. The most-widely used cubic SrTiO$_3$ is chosen as the substrate, which's lattice constant is $3.905$ {\AA}. Although the LaMnO$_3$ bulk is A-type antiferromagnetic, the LaMnO$_3$ thin film on SrTiO$_3$ can become ferromagnetic easily, as repeatedly confirmed in experiments.\cite{Bhattacharya:Prl,Adamo:Apl,Gibert:Nm,Zhai:Nc14} Strain may play an important role in and tiny nonstoichiometry may also contribute to the thin-film ferromagnetism.\cite{Dong:Prb08.2,Hou:Prb14} Such a ferromagnetic background provides a ferroic-order to be controllable by external magnetic field.

\section{DFT calculations}
Our DFT calculations were performed based on the generalized gradient approximation (GGA) with Perdew-Burke-Ernzerhof (PBE) potentials, as implemented in the Vienna \textit{ab initio} Simulation Package (VASP).\cite{Kresse:Prb,Kresse:Prb96} The cutoff energy of plane-wave is $550$ eV. Using the Dudarev implementation,\cite{Dudarev:Prb} the Hubbard repulsion $U_{\rm eff}$ ($=U-J$) for Mn's $3d$ orbitals is tuned from $0$ to $4$ eV. According to previous literature, $U_{\rm eff}=2$ eV on Mn ions is proper to reproduce the experimental properties of LaMnO$_3$, including the Jahn-Teller (JT) and tilting distortions.\cite{Hashimoto:Prb} A large $U_{\rm eff}=8$ eV is imposed on La's $4f$ orbitals to shift the $4f$ states of La away from the Fermi energy.\cite{Dong:Jap14,Hashimoto:Prb} As expected, the calculated band gap for bulk LaScO$_3$ is $4.1$ eV, which agrees with previous calculations.\cite{Ravindran:Jcg,He:Prb12} This band gap, although smaller than the experimental one due to the well-known drawback of DFT, can still lead to a strong confinement of $e_{\rm g}$ electrons of Mn.

To demonstrate an elegant physics, the ideal cubic perovskite structure is tested first (Fig.~\ref{F1}(a)), and the realistic distorted lattice (Fig.~\ref{F1}(c)) will be studied later. For the superlattice based on cubic perovskite, a $12\times12\times5$ Monkhorst-Pack \textit{k}-point mesh centered at $\varGamma$ point is adopted for the Brillouin-zone integrations. While for the distorted lattice, the superlattice is built from the original orthorhombic lattice, then both the lattice constant along the pseudo-cubic ($111$) axis and inner atomic positions are fully relaxed. Since the primary cell is doubled in the distorted case, a $9\times5\times3$ Monkhorst-Pack \textit{k}-point mesh centered at $\varGamma$ point is adopted according to its lattice geometry.

\subsection{Cubic perovskite}
\begin{figure}
\centering
\includegraphics[width=0.5\textwidth]{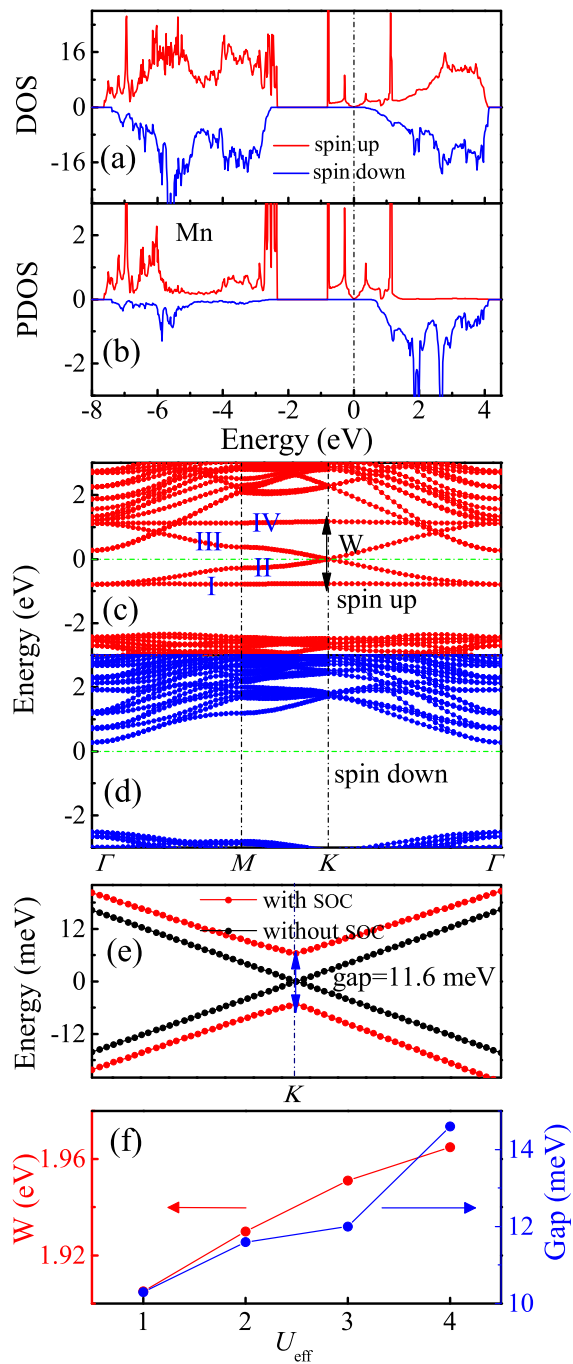}
\caption{(Color online) Electronic structure of the superlattice built by cubic perovskites. (a-e) The DOS and band structures of the (LaMnO$_3$)$_2$/(LaScO$_3$)$_4$ SL for $U_{\rm eff}=2$ eV: (a) Total DOS. The Fermi energy is located at zero; (b) The projected density of states (PDOS) of Mn. (c) The band structure of the spin-up channel in the hexagonal Brillouin zone. $W$ denotes to the bandwidth of bands (I-IV) formed by the $e_{\rm g}$ orbitals; (d) The band structure of the spin-down channel; (e) The band structures magnified around the high-symmetry $K$ point with/without the SOC. (f) The bandwidth $W$ of $e_{\rm g}$ spin-up bands (left axis) and band gap (right axis) with the SOC as a function of $U_{\rm eff}$.}
\label{F3}
\end{figure}

As the first step, the [$111$]-orientated (LaMnO$_3$)$_2$/(LaScO$_3$)$_4$ superlattice based on the cubic perovskite structure is calculated. The typical density of states (DOS) at $U_{\rm eff}=2$ eV is shown in Fig.~\ref{F3}(a).

First, according to the total DOS, the system is a half-metal, namely only the spin-up channel appears around the Fermi level, while there is a large energy gap for the spin-down channel. The states around the Fermi level come from Mn's $3d$ orbitals, as shown in Fig.~\ref{F3}(b). This half-metal behavior is under expectation for manganites and confirmed in all our GGA+$U$ calculation ($U_{\rm eff}=1$, $2$, $3$, $4$ eV as tested).

Second, the system is a semi-metal, namely the DOS at the Fermi level approaches to zero, as in graphene. Such a semi-metal behavior associates with the linear band crossing, i.e. the Dirac-cone-type bands, which are indeed confirmed in our band structure calculation. According to Fig.~\ref{F3}(c), there are four spin-up bands (formed by the $e_{\rm g}$ orbitals) around the Fermi level, as indexed from I to IV according to their energies. The band II and band III touch at the $K$ point, giving the Dirac point, as in graphene. Another interesting character is the prominent flatness of bands I and IV. A quadratic band touching between bands I and II (also between bands III and IV) occurs at the $\varGamma$ point.

In contrast, no spin-down band exists around the Fermi level, in consistent with the DOS. For other values of $U_{\rm eff}$'s from $1$ to $4$ eV, all these characters of DOS's and band structures (not shown) are very similar to the $U_{\rm eff}=2$ eV case, suggesting a robust physical conclusion. The bandwidth from band I to band IV is summarized in Fig.~\ref{F3}(f), which is about $1.9$ eV and slightly increases with $U_{\rm eff}$. Protected by the trigonal symmetry, the Dirac point is robust, which persists even if a very large Hubbard $U_{\rm eff}=8$ eV is applied on Mn's $3d$ orbitals (not shown).

Then, the SOC is enabled in the DFT calculation to examine its effect to band structure. Due to the weak SOC of $3d$ electrons, especially for the $e_{\rm g}$ orbitals, the bands I-IV are almost unaffected in whole. However, a magnified view around the $K$ point shows a gap at the original Dirac point, as shown in Fig.~\ref{F3}(e). According to previous experience,\cite{Xiao:Nc11} such a SOC-opened gap around the Dirac point is very possible to generate topological nontrivial bands. For $U_{\rm eff}=2$ eV, the gap is about $11.6$ meV, equivalent to $\sim100$ K. The $U_{\rm eff}$-dependent gap in the SOC-enabled calculation is shown in Fig.~\ref{F3}(f), which slightly increases with $U_{\rm eff}$. Despite the gap opening, two characters of bands II and III are kept: a) the band dispersion of bands II and III near the $K$ point remain to be linear; b) the top of band II and the bottom of band III remain at the $K$ point.

\subsection{Perovskite with distortions}
\begin{figure}
\centering
\includegraphics[width=0.5\textwidth]{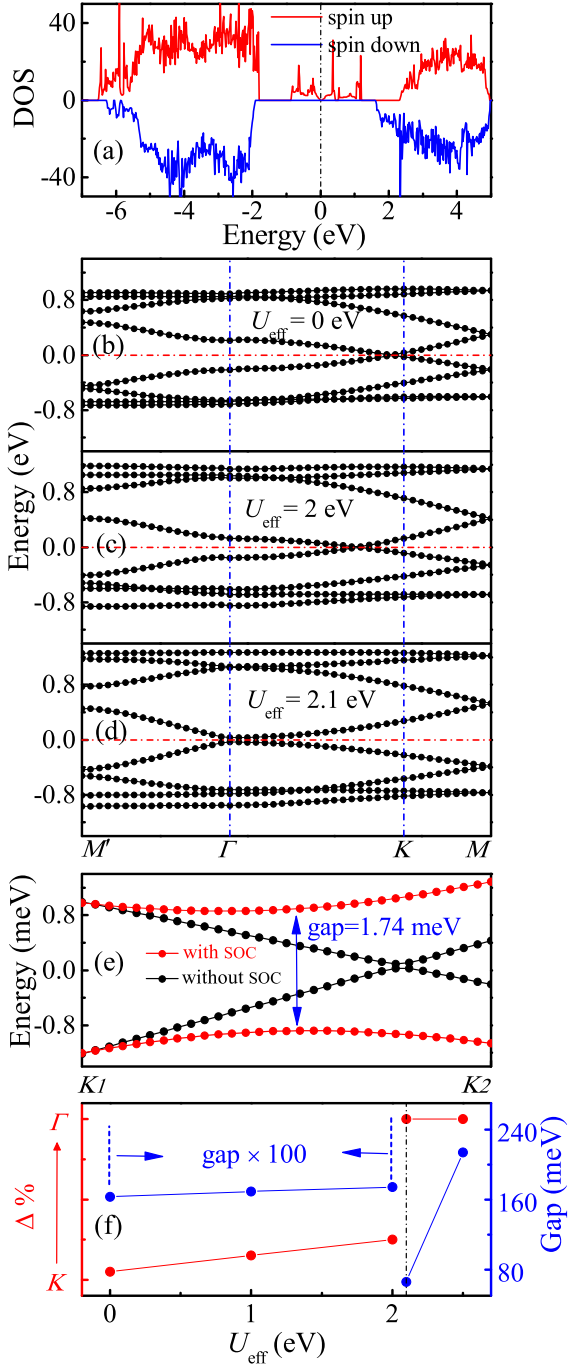}
\caption{(Color online) Electronic structure of the superlattice built by distorted perovskites. (a) Total DOS for $U_{\rm eff}=2$ eV. The Fermi energy is positioned at zero. (b-d) The band structures along a high-symmetric path $M'$-$\varGamma$-$K$-$M$ for different Hubbard $U_{\rm eff}$ values: (b) $U_{\rm eff}=0$ eV; (c) $U_{\rm eff}=2$ eV; (d) $U_{\rm eff}=2.1$ eV. (e) The band structure magnified near the Dirac point with/without the SOC for $U_{\rm eff}=2$ eV. $K_1$ and $K_2$ corresponds ($0.242573$, $0$ ,$0$) to ($0.243905$, $0$, $0$) respectively. (f) The shift of Dirac point from the $K$ point to $\varGamma$ point (left axis) and the band gap (right axis) as a function of $U_{\rm eff}$. Here, the gap for small $U_{\rm eff}$'s ($\leq$ $2$ eV) is opened by the SOC (multiplied by $100$ for better view), while the gap for large $U_{\rm eff}$'s ($>$ $2$ eV) is opened by the Coulomb repulsion.}
\label{F4}
\end{figure}

Subsequently, the calculation is done for the superlattice based on the orthorhombic perovskite structure, which is the realistic structure of LaMnO$_3$ and LaScO$_3$. The DOS at $U_{\rm eff}=2$ eV is shown in Fig.~\ref{F4}(a). Similar to the cubic perovskite structure, the system remains a half-metal and semi-metal, namely the DOS at the Fermi level is fully polarized and approaches to zero. For direct comparison, the $K$ and $M$ points in Fig.~\ref{F4} denote the corresponding points in the Brillouin zone of simple honeycomb as used in Sec.II.A.

Despite these similarities, two nontrivial characters appear in both the GGA and GGA+$U$ calculations on the distorted lattice, as shown in Fig.~\ref{F4}(b-d) and ~\ref{F4}(f). First, the Dirac point moves away from the $K$ point to $\varGamma$ point gradually with increasing Hubbard $U_{\rm eff}$. Second, the Dirac cone opens a gap when $U_{\rm eff}$ is larger than $2$ eV even without SOC. For example, the band gap at the Fermi level is about $65.6$ meV when $U_{\rm eff}=2.1$ eV, which increases to $213.7$ meV when $U_{\rm eff}=2.5$ eV, rendering a Mott insulator.

For those $U_{\rm eff}$'s below the critical value, the band structures are recalculated with SOC enabled. As shown in Fig.~\ref{F4}(e), for $U_{\rm eff}=2$ eV, the Dirac point is opened by SOC, although the band gap ($\sim$$1.7$ meV) is much smaller than the above value ($11.6$ meV at $U_{\rm eff}=2$ eV) of the cubic perovskite structure. The SOC-opened gap increases slightly with $U_{\rm eff}$, as shown in Fig.~\ref{F4}(f). Another character of distorted lattice is that with SOC the band dispersion is no longer linear at the top of valence band and the bottom of conduction band.

Considering the different mechanisms for band gap opening (the SOC-driven one when $U_{\rm eff}\leq2$ eV \textit{vs.} the Hubbard $U$-driven one when $U_{\rm eff}>2$ eV), there must be a sharp turn of gap's value around the critical point, which gives a sudden `jump' for discrete values of $U_{\rm eff}$.

\section{Tight-binding model}
To better understand the physics, a two-orbital tight-binding model for the isolated LaMnO$_3$ bilayer is studied.
In particular, the topological properties, e.g. Berry curvatures and Chern numbers of bands, can be clearly illustrated in the model study.

In the past decades, the two-orbital double-exchange model has been repeatedly confirmed to be very successful to describe the many key physical properties of manganites.\cite{Dagotto:Prp,Dagotto:Bok,Hotta:Rpp} In the current situation, due to the ferromagnetic background, the double-exchange model degenerates into the plain two-orbital tight-binding model. The orbitals involved are two $e_{\rm g}$ ones: $d_{3z^2-r^2}$ and $d_{x^2-y^2}$. Due to the strong Hund coupling, only the spin-up channel is taken into account while the spin-down channel is discarded. The model Hamiltonian reads as:
\begin{equation}
H_{0}=-\sum_{rr'}(t_{rr'}^{ab}d_{r,a}^{\dag}d_{r',b}+h.c.)
\end{equation}
Here, \textbf{r} and \textbf{r}$'$ denote the lattice sites of Mn; $a$, $b$ denote $e_{\rm g}$ orbitals. The hopping amplitude  depends on the orbitals and hopping direction, given by the Slater-Koster formula as follows:\cite{Dagotto:Bok,Dagotto:Prp}
\begin{eqnarray}
\nonumber t_{r,{r}\pm{x}}=\frac{t}{4}\left(
  \begin{array}{cc}
    1 & -\sqrt{3} \\
    -\sqrt{3} & 3 \\
  \end{array}
\right)\\
t_{r,{r}\pm{y}}=\frac{t}{4}\left(
\begin{array}{cc}
1 & \sqrt{3} \\
\sqrt{3} & 3 \\
\end{array}
\right)\\
\nonumber t_{r,{r}\pm{z}}=t\left(
\begin{array}{cc}
1 & 0 \\
0 & 0 \\
\end{array}
\right)
\end{eqnarray}
where $x$, $y$, and $z$ are the unit vectors along three nearest-neighbor Mn-Mn directions, respectively. And the coefficient $t$ is about $0.5$ eV for LaMnO$_3$.\cite{Franchini:Jpcm}

For the $e_{\rm g}$ sector, the spin-orbit coupling (SOC) is absent in the cubic symmetry, since the angular momentum is quenched. However, in ($111$) bilayers, the local symmetry is lowered to $C_3$, therefore, the virtual hopping between $t_{\rm 2g}$ and $e_{\rm g}$ orbitals can give the effective SOC between two $e_{\rm g}$ orbitals. The Hamiltonian for the effective SOC can be written as:\cite{Xiao:Nc11}
\begin{equation}
H_{\rm soc}=-\lambda\sum_{r}(id_{r,a}^{\dag}d_{r,b}+h.c.)
\end{equation}
where $\lambda$ is the effective SOC coefficient.

\begin{figure}
\centering
\includegraphics[width=0.4\textwidth]{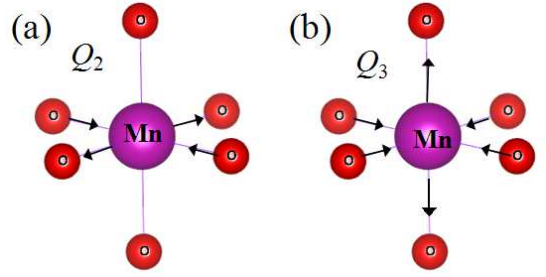}
\caption{(Color online) (a-b) Two Jahn-Teller distortion modes (the black arrows denote the moving of oxygens along Mn-O-Mn bonds): (a) $Q_2$; (b) $Q_3$.}
\label{F2}
\end{figure}

In addition, the Hamiltonian for Jahn-Teller distortion can be written as:\cite{Dagotto:Bok}
\begin{equation}
H_{\rm JT}=\omega\sum_{i}(Q_{2,i}\tau_{x,i}+Q_{3,i}\tau_{z,i})
\end{equation}
where $\omega$ is the electron-lattice coupling coefficient. $Q_2$ and $Q_3$ (Fig.~\ref{F2}(a-b)) are Jahn-Teller distortion modes. $\tau$ is the orbital pseudospin operator, given by $\tau_{x}=d_{b}^{\dag}d_{a}+d_{a}^{\dag}d_{b}$ and $\tau_{z}=d_{b}^{\dag}d_{b}-d_{a}^{\dag}d_{a}$.

Then, the intrinsic anomalous Hall effect conductance is calculated using the standard Kubo formula based on the linear response theory:\cite{Nagaosa:Rmp10,Xiao:Rmp10,Chen:Prb10,Yao:Prl04}
\begin{equation}
\sigma_{xy}^{n}=\frac{e^{2}}{h}\frac{1}{2\pi{i}}\sum_{nmk}\frac{<nk|J_{x}|mk><mk|J_{y}|nk>-H.c.}{[\varepsilon_{n}(k)-\varepsilon_{m}(k)]^{2}}
\end{equation}
Where $|nk>$ is the occupied state and $|mk>$ is the empty state. $J_{x}$ ($J_{y}$) is the $x$ ($y$) components of the current operator. The Berry curvature of each band can be also calculated using the very similar equation, and then the Chern number of each band can be obtained by integrating the Berry curvature over the $1$st Brillouin zone.\cite{Nagaosa:Rmp10,Xiao:Rmp10}

\subsection{Cubic perovskite}
\begin{figure}
\centering
\includegraphics[width=0.5\textwidth]{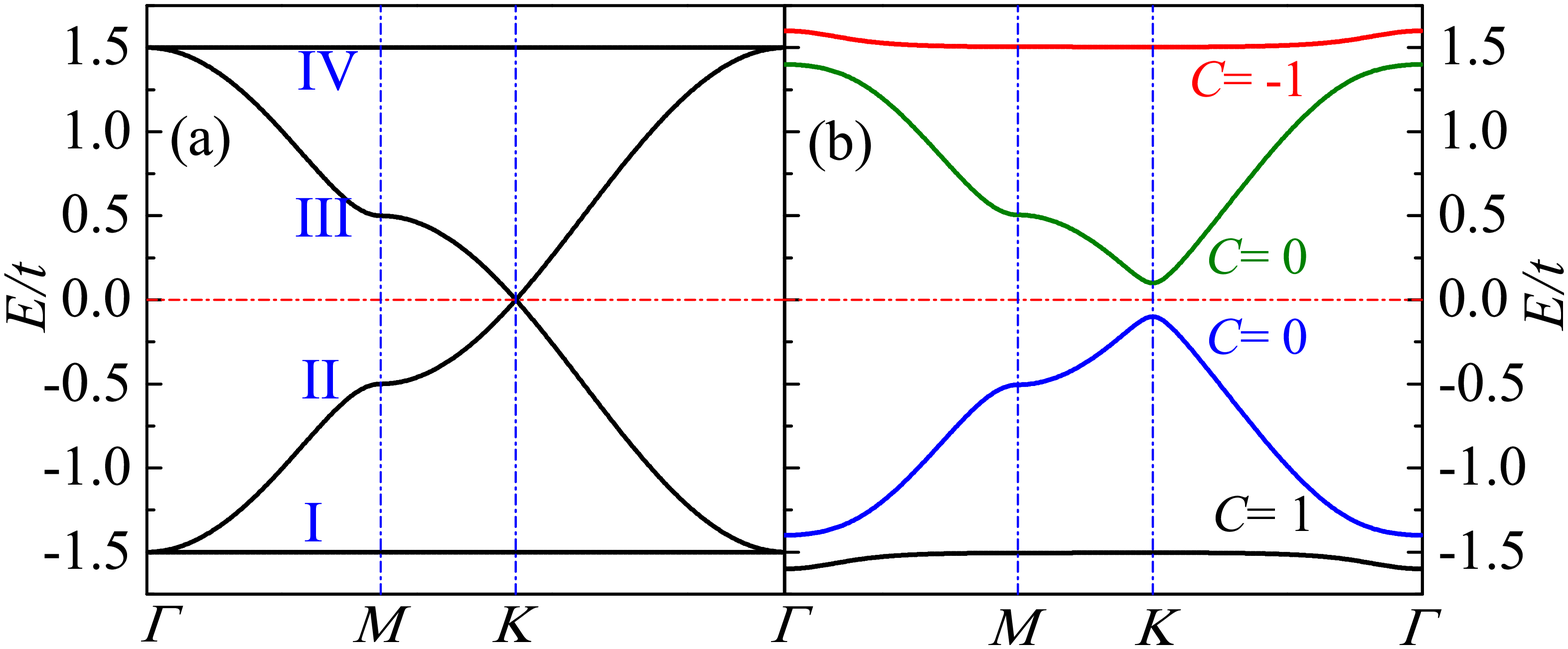}
\caption{(Color online) Band structure of two-orbital model for honeycomb lattice without distortions. (a) Without SOC. (b) With SOC. The Chern number for each band is also shown. Here a large $\lambda=0.1t$ is adopted to make the gaps visible, while the Chern numbers will not be altered by this value.}
\label{F5}
\end{figure}

As in above DFT studies, the model study also starts from the ideal perovskite structure without distortions, as sketched in Fig.~\ref{F1}(d). Without SOC, the band structure of two-orbital tight-binding model, as shown in Fig.~\ref{F5}(a), includes two flat bands (I and IV) and the Dirac cone at the $K$ point formed by bands II and III. In addition, bands I and II (III and IV) form quadratic touching at the $\varGamma$ point. All these features are identical to the corresponding DFT result (Fig.~\ref{F3}(c)), implying the correct physics captured in our two-orbital model.

Then, the SOC is taken into account. As expected, band gaps appear at both the $\varGamma$ point and $K$ point, as shown in Fig.~\ref{F5}(b). Analytically, the band gap equals $2\lambda$ at the $K$ point and $\varGamma$ point.

In addition, the topological index, e.g. the Chern number, is calculated for the SOC-enabled bands by integrating the Berry curvature of each band, giving ($1$, $0$ , $0$, $-1$) as shown in Fig.~\ref{F5}(b). Such a topological nontrivial band structure can give rise to the quantized anomalous Hall effect when the system is ideally stoichiometric.\cite{Nagaosa:Rmp10,Xiao:Rmp10}

\begin{figure}
\centering
\includegraphics[width=0.5\textwidth]{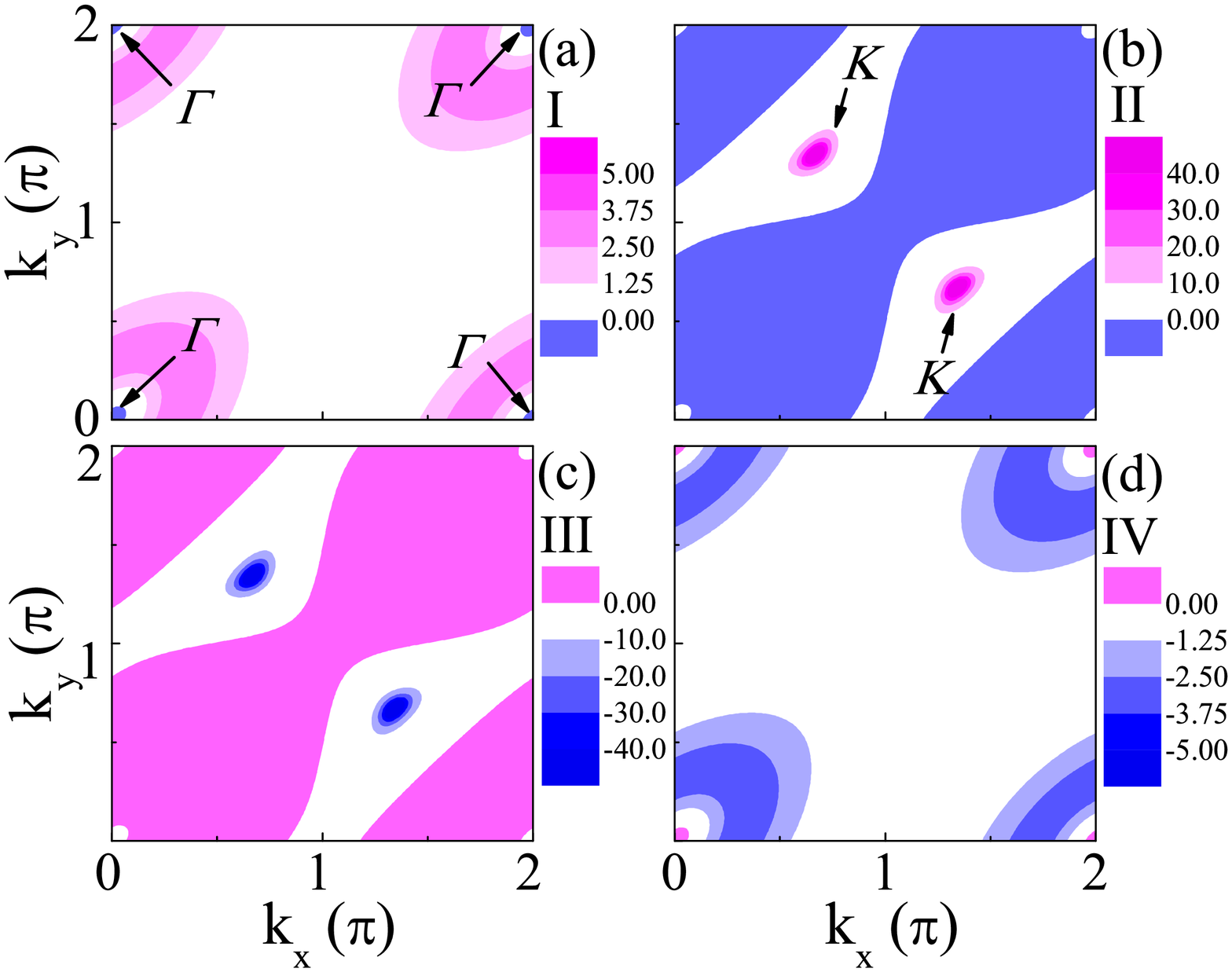}
\includegraphics[width=0.4\textwidth]{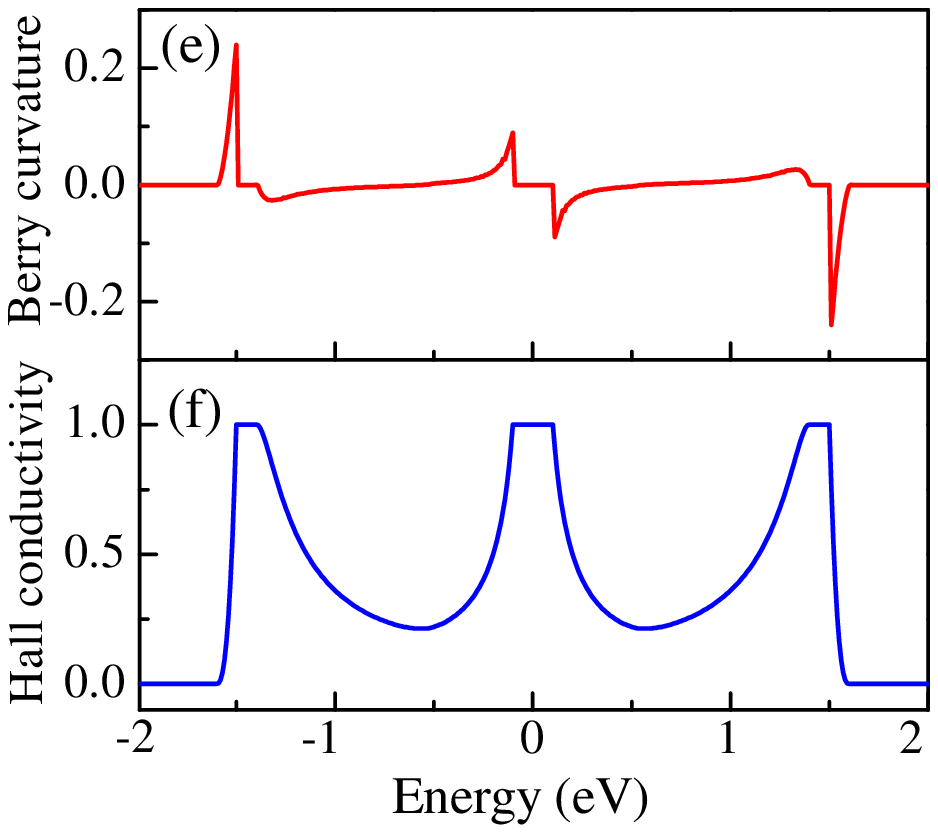}
\caption{(Color online) (a-d) The distribution of Berry curvature in the $1$st Brillouin zone for bands I to IV, respectively. Noting the color schemes are identical in (a) and (d) (also (b) and (c)). (e) The energy-dependent Berry curvature. (f) The intrinsic anomalous hall conductance in unit of $e^{2}/h$ as a function of chemical potential. (f) is obtained by integrating the Berry curvature (e) over the energy.}
\label{F6}
\end{figure}

The Berry curvature of each band is shown in Fig.~\ref{F6}, whose integral over the $1$st Brillouin zone is the Chern numbers. It is clear that the distribution of Berry curvature is nontrivial for all four bands. Even for the bands II and III whose Chern numbers are zero, the Berry curvatures have peaks at the $K$ points, implying a $1/2$ Chern number contributed by each Dirac cone in the SOC-enabled case.\cite{Nagaosa:Rmp10,Xiao:Rmp10} Then for band II (or band III), the background of Berry curvature cancels the contribution of two Dirac cones, giving a zero Chern number. In this sense, although the Chern numbers for bands II and III are zero, they are topological nontrivial when partially filled. The energy-dependent Berry curvature is shown in Fig.~\ref{F6}(e). The intrinsic anomalous hall conductance as a function of chemical potential is shown in Fig.~\ref{F6}(f).

\subsection{Perovskite with distortions}
Finally, the distorted orthorhombic system (see Fig.~\ref{F1}(e)), is studied using the tight-binding model. For the distorted lattice, an important physical issue is the Jahn-Teller distortion which is quite prominent in LaMnO$_3$ bulk and contributes to the special $3x^{2}-r^{2}$/$3y^{2}-r^{2}$ type orbital ordering.

In fact, it is well accepted that the typical Mottness of two $e_{\rm g}$ orbitals in manganites involves the charge ordering and orbital ordering, which originate from the distortions of structure.\cite{Dagotto:Bok,Dagotto:Prp,Tokura:Sci,Hotta:Rpp,Tokura:Rpp} According to substantial literature, the model with either purely Coulomb interactions or Jahn-Teller electron-lattice coupling can be successful to reproduce many experimental properties.\cite{Dagotto:Bok,Dagotto:Prp,Hotta:Rpp} The physical similarity is that both interactions split the multi-$e_{\rm g}$-bands, giving rise to the Mottness. In this sense, the Jahn-Teller effect and Coulomb repulsion are qualitatively equivalent, although not exactly identical. Therefore, the topological-Mott transition obtained in above DFT calculation, can be qualitatively mimicked by the Jahn-Teller interaction in the tight-binding model.

\begin{figure}
\centering
\includegraphics[width=0.4\textwidth]{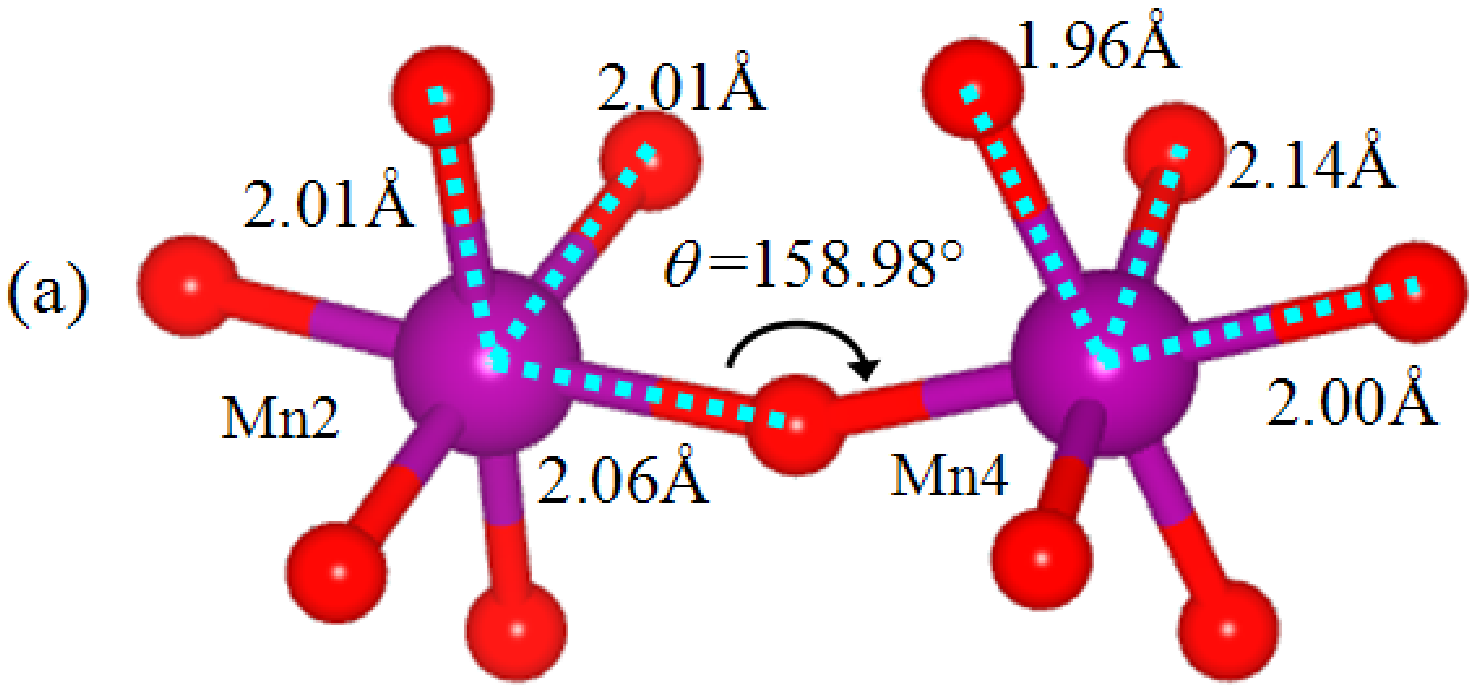}
\includegraphics[width=0.5\textwidth]{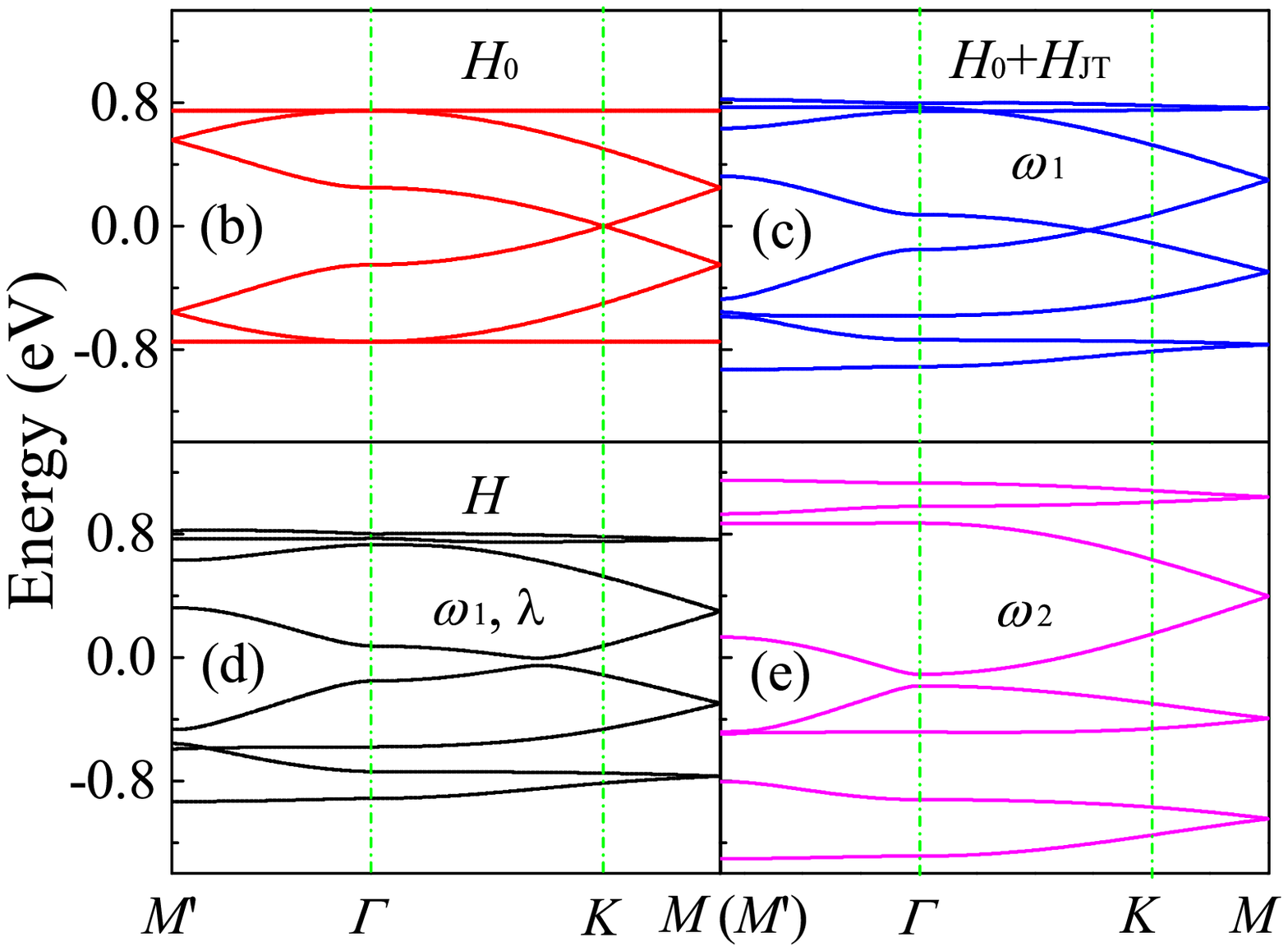}
\caption{(Color online) (a) The sketch of optimized Mn-O-Mn bond angles ($\theta$) and Mn-O bond lengths for the orthorhombic system. (b-e) The tight-binding band structures for the expanded lattice (Fig.~\ref{F1}(c)): (b) Without the Jahn-Teller distortion; (c) With the Jahn-Teller distortion; $\omega_{1}$=$1.3$ eV/{\AA}; (d) With the Jahn-Teller distortion ($\omega_{1}$=$1.3$ eV/{\AA}) and SOC. Here an overlarge $\lambda$=$30$ meV is used to make the band gaps visible. (e) The Mott transition is generated by a larger Jahn-Teller coefficient ($\omega_{2}$=$2.7$ eV/{\AA}) without SOC.}
\label{F7}
\end{figure}

By measuring the DFT-relaxed Mn-O bonds (see Fig.~\ref{F7} (a) for example), the Jahn-Teller modes are obtained. For example, when $U_{\rm eff}$=$2$ eV, $Q_2$(Mn1)=$0.303$ {\AA}, $Q_3$(Mn1)=$-0.067$ {\AA}, $Q_2$(Mn2)=$-0.018$ {\AA}, $Q_3$(Mn2)=$0.073$ {\AA}, $Q_2$(Mn3)=$0.018$ {\AA}, $Q_3$(Mn3)=$0.073$ {\AA}, $Q_2$(Mn4)=$-0.303$ {\AA} and $Q_3$(Mn4)=$-0.067$ {\AA}. By fitting the DFT bands (Fig.~\ref{F4}(c)) at the $\varGamma$ point, the double exchange hopping amplitude $t$ and the Jahn-Teller coefficient $\omega$ can be obtained as $0.45$ eV and $1.3$ eV/{\AA} respectively.\footnote{The fitted value of $\omega$ increases with the value of $U_{\rm eff}$ used in the DFT calculation.}

The model's band structures for the expanded unit cell without and with the Jahn-Teller distortions are shown in Fig.~\ref{F7}(b) and ~\ref{F7}(c), respectively. When the Jahn-Teller interaction is not considered (by setting $\omega=0$), Fig.~\ref{F7}(b) shows the Dirac point locating at the $K$ point, as expected. And the top and bottom flat bands are double-degenerated due to the Brillouin zone folding since the unit cell (Fig.~\ref{F1}(e)) is doubled. When the Jahn-Teller interaction is considered, the Dirac point moves to the $\varGamma$ point, and the degeneration of top and bottom bands is split, as shown in Fig.~\ref{F7}(c). All these features have also been observed in above DFT calculations, implying the correct physics captured in the model. As shown in Fig.~\ref{F7}(d), a gap is opened by incorporating the SOC, which is also identical to above DFT result presented in Sec.II.B. Moreover, the Mott transition can be obtained by using a Jahn-Teller coefficients (see Fig.~\ref{F7}(e)).

\begin{figure}
\centering
\includegraphics[width=0.4\textwidth]{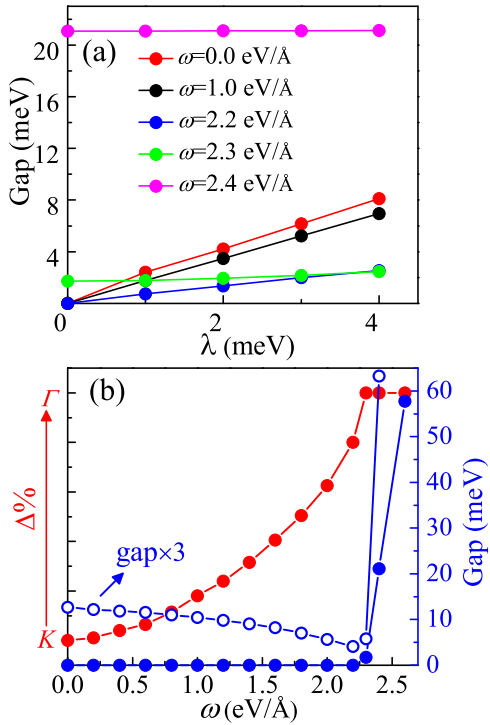}
\caption{(Color online) Results of the two-orbital model with Jahn-Teller distortion and SOC. (a) The band gap as a function of the effective SOC parameter $\lambda$ with various Jahn-Teller coefficients. (b) The shift of Dirac point (red circles) from the $K$ point to $\varGamma$ point (left axis) as a function of the Jahn-Teller coefficient without SOC. The band gap (right axis) as a function of the Jahn-Teller coefficient without SOC (blue solid circles) and with SOC (multiplied by $3$ for better view) (blue open circles).}
\label{F8}
\end{figure}

Thus, to better understand underlying mechanism, the band gap around the Fermi level is studied upon the tuning of relevant parameters of Eq. 3 and Eq. 4 (e.g. $\lambda$ and $\omega$), as shown in Fig.~\ref{F8}(a-b). For simplify, the aforementioned values of $Q$'s at $U_{\rm eff}$=$2$ eV will be adopted. First, for a small Jahn-Teller coefficient (e.g. $\omega\leq2.2$ eV/{\AA}), the band gap is linearly in proportional to the SOC coefficient $\lambda$. In this sense, in this small $\omega$ region, the Jahn-Teller distortion will not alter the topological feature of bands, although the position of Dirac cone moves to $\varGamma$ point with increasing $\omega$ (Fig.~\ref{F8}(b)). When $\omega$ goes beyond this criterion, e.g. $\omega=2.3$ eV/{\AA}, the Jahn-Teller interaction itself opens a band gap while the SOC can only slightly tune the gap¡¯s value. Therefore, such a band gap suggests a Mott insulator, as in the LaMnO$_3$ bulk.\cite{Dong:Prb08.3} In this sense, the Jahn-Teller distortion, which prefers the Mottness, can suppress the topological phase of LaMnO$_3$ bilayer. In fact, even in the small $\omega$ region, the value of SOC-opened band gap is suppressed with increasing $\omega$.

\section{Conclusion}
In summary, using the first-principles calculation and tight-binding model, we studied the electronic structures of the ($111$) LaMnO$_3$ bilayer sandwiched in LaScO$_3$ barriers. In general, the system presents half-metal and semi-metal (with Dirac cones) behavior when the spin-orbit coupling is not taken into consideration. Then, the spin-orbit coupling can open a gap for the Dirac cone, rendering a topological magnetic insulating phase (Chern insulator), as confirmed by the tight-bonding model. In addition, the orthorhombic distortion, can move the Dirac cone  position from the high symmetric $K$ point to $\varGamma$ point, driven by the Jahn-Teller effect as well as the Coulomb repulsion. A Mott gap appears when the Jahn-Teller effect (or the equivalent Coulomb repulsion) is strong enough, suppressing the topological properties of bands. Thus, to pursuit a magnetic topological phase in perovskite ($111$) bilayer, it is better to suppress the lattice distortions. Among all undoped $R$MnO$_3$ ($R$: rare earth), LaMnO$_3$ owns the weakest lattice distortion, implying the best choice. Furthermore, other special manganites, e.g. quadruple perovskites with weak Jahn-Teller modes,\cite{Zhang:Prb11} and proper strain from substrates, may provide more feasible choices to obtain perovskite structure that more close to the idea cubic limit.

\acknowledgments{S.D. was supported by the NSFC (Grant Nos. 11274060 and 51322206) and Jiangsu key laboratory for advanced metallic materials (Grant No. BM2007204). Y.Y. was supported by the MOST Project of China (Grant Nos. 2014CB920903, 2011CBA00100), NSFC (Grant Nos. 11174337 and 11225418), Specialized Research Fund for the Doctoral Program of Higher Education of China (Grant No. 20121101110046). Y.K.W. was supported by the Jiangsu Innovation Projects for Graduate Student (Grant No. KYLX15\underline{ }0112).}

\bibliographystyle{apsrev4-1}
\bibliography{ref}
\end{document}